\documentclass[twocolumn,secnumarabic,amssymb,amsmath,nofootinbib,tightenlines,nobibnotes,aps,prb,showpacs]{revtex4}
\usepackage{graphicx}

\begin{document}

%\draft
%\showrefnames
\date{\today}

\title{ From Bubble to Skyrmion:
Dynamic Transformation Mediated by a Strong Magnetic Tip }
\author{A.S. Kirakosyan$^1$}
%\address
\affiliation{Department of Physics, Texas A\&M University,
College Station, Texas 77843-4242}
\author{V.L.Pokrovsky$^{1,2}$}
%\address
\affiliation{$^1$Department of Physics, Texas A\&M University, College
Station, Texas 77843-4242\\
$^2$Landau Institute of Theoretical Physics, Moscow, Russia}

\begin{abstract}
Skyrmions in thin metallic ferromagnetic films are stable due to
competition between the RKKY interaction and uniaxial magnetic
anisotropy. We mimic the RKKY interaction by the
next-nearest-neighbors ferromagnetic and antiferromagnetic
exchange interactions. We demonstrate analytically and numerically
dissipative transformation of a bubble created by a strong
cylindrical magnetic tip into a stable Skyrmion.
\end{abstract}
\maketitle

\section{Introduction}
\label{intro}
Recently a significant  progress has been achieved
in the nanofabrication and design of new technologies for very
high density storage on hard disks (densities above 500
Gbit/in$^{2}$ )   \cite{kirk&all00} and magnetoresistive random
access memory (MRAM).  For the further miniaturization of the data
storage and the magnetic memory it is important to decrease the
size of nanoelements. Magnetic defects in such nanoelements
storing bits of information can have very small size. The
properties of Skyrmion in thin metallic magnetic films (the small
size about 1 nm and stability)  may allow to use their
two-dimensional (2D) arrays in magnetic storage  devices. To the
best of our knowledge a magnetic memory based on Skyrmions was not
developed yet. There are no analogs of storage of so extremely
high density.

Skyrmions are strongly localized scale invariant analogs of magnetic
bubble domains (MBD) \cite{bubble}. The axially  symmetric Skyrmion
solution for boson fields minimizing  the Hamiltonian and describing  the
structure of nuclear matter was first found by Skyrme \cite{Skyrm58}.
Belavin and Polyakov based on nonlinear $\sigma$-model have established
that Skyrmion is a topologically nontrivial metastable minimum of energy
\cite{BP75}. Since then the properties of Skyrmions in two-dimensional
magnets and thin magnetic films have been widely investigated in
last few decades
\cite{voronov&all83,ivanov&all90,Abanov&Pokrovsky98,abdullaev&all99,sheka&all01}.

Slow relaxation spin dynamics of Skyrmion was studied by Abanov
and Pokrovsky \cite{Abanov&Pokrovsky98}. They calculated the
radius of the Skyrmion as a function of time. They have found that
the interaction of the Ruderman-Kittel-Kasuya-Yosida (RKKY) type
together with the uniaxial magnetic anisotropy can stabilize the
Skyrmion.

In quasi-2D antiferromagnets the Skyrmions were indirectly
observed by F. Waldner by the ESR \cite{waldner83} and the heat
capacity measurements \cite{waldner86} in a good agreement with
the Belavin and Polyakov (BP) theory. The appearance of Skyrmions
in the QHE was predicted by Sondhi et al. \cite{sondhi&al93}. In
the regime of fractional QHE Skyrmions were discovered
experimentally in GaAs quantum wells by means of optically pumped
NMR and by measurements of the Knight shift, which is proportional
to the electron magnetization \cite{barrett&all95}. For integer
QHE Skyrmions were detected in thermal magnetotransport
measurements \cite{schmeller95} and in optical absorption
\cite{aifer96}.

Recently another topological excitations, magnetic vortices were
observed and studied experimentally in permalloy and Co nanodisks
by means of the Lorentz microscopy and Magnetic Force Microscopy
(MFM) \cite{schneider&all00}.  The evolution and stability of such
magnetic vortices in a small isotropic cylindrical magnetic dot
was analyzed theoretically  in \cite{guslienko&metlov01}.

The equilibrium size of Skyrmion in  2D thin magnetic films is
extremely small. This is one of serious obstacles for its
experimental observation. For soft magnets the size of vortices is
defined by competition between the exchange and magnetic dipole
interaction. In uniaxial magnets, strong uniaxial anisotropy
together with the fourth order exchange interaction leads to small
size of Skyrmion. Recent great progress in nanotechnology gives a
chance to  detect experimentally Skyrmions in thin magnetic films.
The experimental methods employed for studying Skyrmions in QHE
systems and vortices in soft magnetic dots can be  also applied to
2D uniaxial magnets. So far the only technique potentially capable
of atomic resolution is the spin-polarized scanning tunneling
microscopy (SPSTM).

Due to stability of the Skyrmion in an uniaxial thin metallic
magnetic film it is possible to create it by a strong magnetic
tip. In this paper we study the properties of the MBD created by
the tip as well as its relaxational dynamics when tip moves away
uniformly and vertically from the film. When the distance between
the film and the tip exceeds a critical value, it is not able to
confine the MBD anymore and the latter collapses and gradually
transforms into a stable Skyrmion. For numerical simulations we
use a stabilization mechanism provided by the interplay between
next nearest neighbor antiferromagnetic interaction and the
magnetic anisotropy. We also show that the demagnetization energy
of the Skyrmion is equivalent to the renormalization of the single
ion anisotropy constant. We find that demagnetizing field produced
by a single Skyrmion is strongly localized around Skyrmion center.

The plan of this paper is as follows. In Section \ref{model} we
introduce a theoretical model and equations of motion. We
calculate the demagnetization energy of the Skyrmion in Section
\ref{demag}. Section \ref{tip} is devoted to the magnetic field
generated by a cylindrical tip. In Section \ref{static} we study
static properties of the MBD and the Skyrmion solving numerically
Landau-Lifshitz equation with the fitting method. In Section
\ref{discrete} we study analytically and numerically a discrete
spin model with the dipolar interaction and find stability
condition for the static Skyrmion. Relaxational dynamics of the
transformation of a MBD into a Skyrmion in the presence of an
uniformly moving tip is considered  in Section \ref{relax}. Our
conclusions and discussions are  presented in the last section
\ref{conclusion}.

\section{Model}
\label{model}

We consider the two-dimensional Heisenberg model with the exchange
interaction up to the fourth order derivatives and uniaxial
anisotropy in continuous approximation. Its Hamiltonian has a
following form
\begin{equation}
{\cal H} = h \int W({\bf M})\, d^2 x
\label{H}
\end{equation}
where
$W({\bf M}) =
W_e({\bf M}) + W_a({\bf M}) + W_Z({\bf M})$ and
\begin{eqnarray}
W_e({\bf M}) &=& {\alpha \over 2} \left((\nabla {\bf M})^2 +
{a^2 \kappa \over 2 \, M_0^2}
(\nabla {\bf M})^4 \right)
\label{W1}\\
W_a({\bf M}) &=& {\beta \over 2} (M_0^2 - M_z^2), \label{W2}\\
W_Z({\bf M}) &=& H_z\,M_0 - \left( {\bf H}(r,d(t)) + {1 \over 2}
{\bf H^{D}} \right)\,{\bf M},\label{W3}
\end{eqnarray}
%%%%%%%%%%%%%%%%%%%%%%%%%%%
%%%%%%%%%%%%%%%% Hamiltonian
%%%%%%%%%%%%%%%%%%%%%%%%%%%
Here the energy  density $W({\bf M})$ consists of the exchange
energy $W_e({\bf M })$, the uniaxial anisotropy energy $W_a({\bf
M})$; $W_Z({\bf M})$ contains the interaction energy of
cylindrical tip with the magnetic film and the demagnetization
energy.
%%%%%%%%%%%%%%%%%%%%%%%%%%%%%%%%%%
%%%%%%%% Dynamical Equations
%%%%%%%%%%%%%%%%%%%%%%%%%%%%%%%%%%
In our notations $\alpha$ is the exchange interaction constant,
$\beta$ is the easy-axis anisotropy constant, $\kappa$ is a
positive dimensionless constant, characterizing the fourth order
exchange interaction (we consider just this term because of its
simplicity), $M_0$ is the spontaneous magnetization of the film,
${\bf H}(r, d)$ is the magnetic field created by the magnetic tip
(cylinder) at the surface of the film as a function of the polar
radius $r$ and the distance $d$ between the cylinder and the plane
of the film; $h$ is the film thickness.  In a thin magnetic film
the demagnetization energy of a localized magnetic excitation with
the radius $R_e$ at $R_e \ll h$ simply renormalizes the anisotropy
constant $\beta$ ($\beta \rightarrow \beta - 4 \pi$). In the
opposite case $R_e > h$ the demagnetization energy is more
complicated. Below we calculate it for the Skyrmion. For iron the
exchange length $l_{ex} = 1.5 nm$. Therefore, an ultrathin iron
film consists of few monolayers.

The dynamics of a classical 2d ferromagnet is governed by the
Landau-Lifshitz equation (LLE) for the magnetization vector ${\bf
M}$

\begin{eqnarray}
\dot{\bf M}({\bf  r},t)&=& \gamma \,{\bf M}({\bf r},t)  \times
\frac{\delta {\cal H} [{\bf M}]}{\delta {\bf M}({\bf r},t)} + \nonumber \\
{\Gamma \,  \dot{\bf M}({\bf  r},t)  \times  {\bf M}({\bf r},t)  \over   M_0
}
\label{LLE}
\end{eqnarray}
here $\gamma = g \mu_B / \hbar$; $g$ and $\mu_B$ are the
gyromagnetic ratio and the Bohr magneton, respectively, and
$\Gamma$ is the dimensionless Gilbert damping constant. The LLE
equation (\ref{LLE}) in angular variables $\theta$ and $\phi$ can
be derived from the Lagrange-Hamilton variational principle with
the following Lagrangian density ${\cal L}(\theta, \phi)$
\begin{equation}
 {\cal L}(\theta, \phi) = {M_0 \over \gamma} \, (1 - \cos \theta) \,
 {\partial \phi \over \partial t} - W(\theta, \phi)
\label{L}
\end{equation}
In the static case the LLE Eq. (\ref{LLE})  read:
\begin{equation}
M_0^2 \,  {\delta {\cal H} \over \delta {\bf M({\bf r})}} =  {\bf M({\bf
r})}\,\left( {\bf M({\bf r})} \cdot {\delta {\cal H} \over \delta {\bf M({\bf
r})}} \right)
\label{stLLE}
\end{equation}
The right-hand side of Eq. (\ref{stLLE}) originates from a
constraint: ${\bf M}^2 = M_0^2$ at any point ${\bf r}$.

The classical ground state is double degenerate due to uniaxial
magnetic symmetry. A topologically nontrivial minimum of the
Hamiltonian (\ref{H}) corresponds to the Skyrmion solution
\begin{eqnarray}
M^{s}_{x} &=&  M_0 \, {2 \, R_s \, r \over R_{s}^2 + r^2} \cos(\varphi + \psi),
\nonumber \\ M^{s}_{y} &=& M_0 \, {2\, R_s \,r \over R_{s}^2 + r^2} \sin(\varphi
+ \psi), \nonumber \\ M^{s}_{z} &=& M_0\,{r^2  - R_{s}^2 \over r^2 + R_{s}^2}.
\label{skyrm}
\end{eqnarray}

The Skyrmion is characterized by its radius $R_s$ and an angle
$\psi$ of the global spin rotation;  $r$ and $\varphi$ are polar
coordinates. For isotropic magnets the energy of Skyrmion does not
depend neither on $R_s$ nor on $\psi$. In real systems the
degeneracy with respect to $R_s$ is lifted by uniaxial magnetic
anisotropy and by the fourth order exchange interaction.

Another static solution of the LLE is the well known MBD, which
was the subject of intensive study during last few decades. Here
we consider an MBD with the topological charge one. The structure
of such a soliton can be approximately described by the
"domain-wall" ansatz:
\begin{eqnarray}
M^{b}_{x} &=& M_0 \,   \mbox{sech}{ r- R_b \over l_0}  \cos(\varphi + \psi),
\nonumber \\ M^{b}_{y} &=& M_0  \,  \mbox{sech}{ r- R_b \over l_0} \sin(\varphi
+ \psi), \nonumber \\ M^{b}_{z} &=& M_0  \, \tanh {r - R_b \over l_0},
\label{MBD}
\end{eqnarray}
where $R_b$ is the MBD radius. Since we consider an ultrathin
magnetic film, the MBD exists only in the presence of an external
magnetic field and collapses without this field.

\section{Demagnetization fields}
\label{demagnet}

\begin{figure}
\begin{center}
%\vspace{15 mm}
\includegraphics[width=0.9\columnwidth,height=0.75\columnwidth]{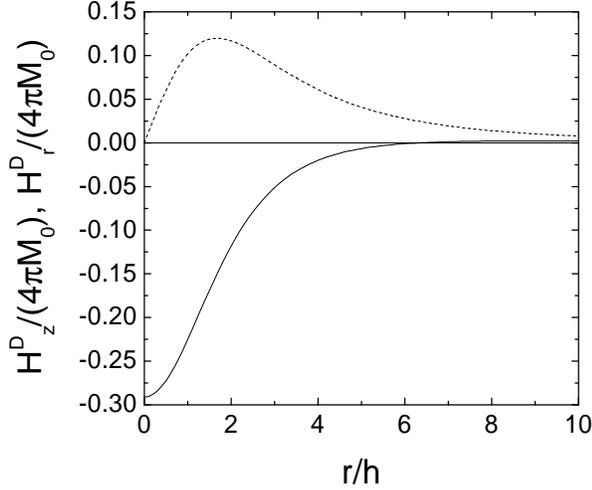}
\caption{ Demagnetizing field of the Skyrmion (\ref{skyrmdemag})
versus polar radius normalized to the film thickness $r/h$. Solid
and dashed lines correspond to $H^D_{z}/(4 \pi M_0)$ and
$H^D_{r}/(4 \pi M_0)$, respectively. The radius of Skyrmion $R_s =
2 h$. %\vspace{5 mm}
} \label{demag}
\end{center}
\end{figure}

Let us now return to the demagnetization energy. The magnetostatic
problem for the non-topological MBD was solved first by Thiele
\cite{thiele69,thiele70}. He expressed $ {\cal H}_D$ in terms of
elliptic integrals. The Thiele's solution for demagnetizing field
$x^{-1} (\partial {\cal E}_D / \partial x) $ can be approximated
with surprisingly good accuracy by  \cite{callen&josephs71}
\begin{equation}
x^{-1} (\partial {\cal E}_D / \partial x) \approx - (1 + 3 x/2)^{-1},
\label{approxhd}
\end{equation}
where dimensionless radius   $x  = R_b/h$ and dimensionless energy
${\cal E}_D  = (16 \pi^2 M_0^2 h^3)^{-1} {\cal H}_D  $ are
introduced. For magnetostatic interaction the topological charge
does not play any role. Hence the Thiele's solution persists when
dipolar forces are taken in account.

For ultrathin magnetic films consisting of few layers the ratio of
the exchange length to the thickness is $l_{ex}/h > 4/3$.
Therefore, the MBD solution is unstable without external magnetic
field.

The reading of the information stored by Skyrmions can be
performed by the measurement of the Skyrmion magnetic field. Below
we calculate this field and its energy. The Fourier transform of
the Skyrmion magnetic field  has the following form
\begin{equation}
 {\bf h}^{D}({\bf k}) = - 4 \pi \,{\bf k} \, {({\bf k} \cdot {\bf m}({\bf k})) \over {\bf k}^2}
\label{demagfield}
\end{equation}
The magnetostatic energy in terms of the magnetization
Fourier-components reads:
\begin{equation}
{\cal E}_D  = {1 \over (2 \pi)^2}  \int {\mid  {\bf k} \cdot {\bf m}({\bf k}) \mid^2
\over {\bf k}^2 } d{\bf k}
\label{demagskyrm}
\end{equation}
To calculate the magnetic field we substitute the Fourier
transform of the Skyrmion magnetization (\ref{skyrm}) into
(\ref{demagfield}) and perform the inverse Fourier transformation.
The result is as follows:
\begin{eqnarray}
 H^{D}_{z} = 4 \pi  M_0 \, R_{s}^2 \int_{0}^{\infty} F_{z}(q,z,h) \, K_{0}(q R_s) \, J_{0}(q r)
 \, q  dq  \nonumber \\
 - 4 \pi M_0 \Theta[h/2 - |z|]
 \nonumber  \\
 H^{D}_{r} = 4 \pi  M_0 \, R_{s}^2 \int_{0}^{\infty} F_{r}(q,z,h) \, K_{0}(q R_s) \, J_{1}(q r)
 \,  q  dq,
\label{skyrmdemag}
\end{eqnarray}
%%%%%%%%ADD ?????????
where
\begin{eqnarray}
 F_{z} &=& \mbox{sign}\left(  {h \over 2}  - z\right)\, e^{- q \left| {h \over 2}  - z \right|} +
 \mbox{sign}\left(  {h \over 2}  + z\right)\, e^{- q \left|  {h \over 2}   + z \right|}
 \nonumber  \\
 F_{r} &=& e^{- q \left| {h \over 2}  - z \right|} - e^{- q \left| {h \over 2} + z \right|},
 \label{FzFr}
\end{eqnarray}
The origin of the frame of reference coincides with the Skyrmion
center. The Skyrmion demagnetizing field for $R_s/h = 2$ is shown
in Fig. \ref{demag}. The demagnetization energy of the Skyrmion
(\ref{demagskyrm})in units $E_0$ has the following form:
 \begin{equation}
  {\cal E}_D = {\cal E}_{D}^{1}  - {4 \pi \over \beta} \, {\cal E}_{an},
\label{ed}
\end{equation}
where the last term is a part of magnetostatic energy
renormalizing anisotropy constant: $\beta \rightarrow \beta -
4\pi$ \cite{ivanov&all90} and ${\cal E}_{D}^{1} $ has the form
\begin{eqnarray}
  {\cal E}_{D}^{1}  &=&
 {\pi \rho^3 \over 8} \left(3 \pi - 4 E \left( m \right) -
  2 K\left( m \right)  \right) \nonumber \\ &&
  - {1 \over 18} \, {_{p}F_{q}} \left( \{1,1,3\},\{{5 / 2},{5 / 2}\}, m \right)
  \nonumber \\ &&
  + {\rho^2 \over 2} \, {_{p}F_{q}} \left( \{1,1,1\},\{{3 / 2},{3 / 2}\}, m \right),
    \label{demagen}
 \end{eqnarray}
where $\rho = R_s/h$, $m = 1/(4 \rho^2)$, $K(m)$ and $E(m)$ are
the complete elliptic integrals of first and second kind,
respectively, and  ${_{p}F_{q}}(\{a_1, \ldots,  a_p\},  (\{b_1,
\ldots,  b_q\}, m)$ is the generalized hypergeometric function. We
consider ultrathin film $\rho \geq 1$. Employing an appropriate
expansions for the elliptic integrals and hypergeometric functions
we find
\begin{equation}
  {\cal E}_{D}^{1}  \approx {\rho^2 \over 2} + {\pi^2 \, \rho \over 128} - {26 \over 225}  -
 {1 \over 150 \rho^2 }; \qquad \rho \gg 1
\label{expanddemag}
\end{equation}
The expression (\ref{expanddemag}) can be  replaced  by $\rho^2/2
+ \pi^2 \, \rho/128 $ in the interval $\rho > 1$ with the accuracy
of half percent.

The spin - polarized scanning tunneling microscopy (SP-STM) and
spectroscopy (SP-STS) offers a great potential to unravel the
structure of a single Skyrmion  and Skyrmion arrays on the
nanometer scale.

\section{Cylindrical magnetic tip}
\label{tip}

\begin{figure}
\begin{center}
\vspace{5 mm}
\includegraphics[width=0.9\columnwidth,height=0.75\columnwidth]{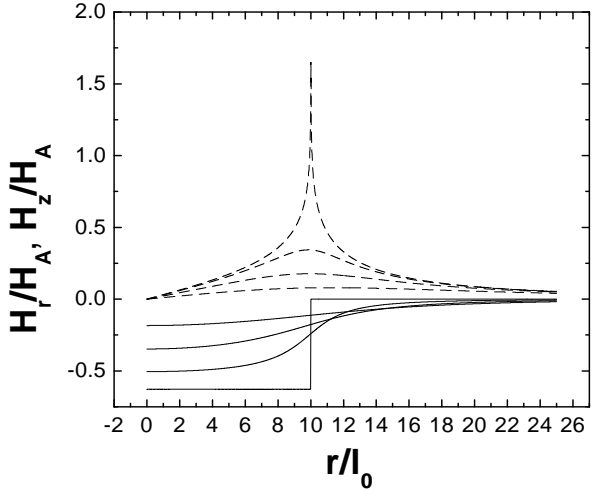}
\vspace{3 mm} \caption{ Components of the magnetic field of the
tip (\ref{hz}) and (\ref{HzHr}) at a distance $d$ from the film:
$d/l_0$ 0, 2, 5 and 10. $H_A = \beta M_0$ is the anisotropy field
of the film and $l_0 = 10\, nm$ is the magnetic length. Solid and
dashed lines correspond to $H_z$ and $H_r$ components,
respectively. The radius of magnetic tip $R = 10 \,l_0 = 100
\,nm$, the ratio of magnetization of the tip to that of the film
is $M_{t}/M_{0} = 10$. } \label{Htip}
\end{center}
\end{figure}

Consider a semi-infinite cylindrical magnetic tip.  The stray
magnetic field of the cylindrical tip is well known. Its $z$ and
$r$ components are
\begin{eqnarray}
\label{hz}
H_z(r,d) &=& - 2 \pi M_t R \int_0^\infty dq\,J_1(q R)\, J_0(q r)\, e^{-q d}, \\
H_r(r,d) &=&  2 \pi M_t R \int_0^\infty dq\, J_1(q R)\, J_1(q r)\, e^{-q d},
\label{HzHr}
\end{eqnarray}
where $M_t$, $R$ and $d$ are the magnetization, the radius of
cylinder and the distance between the tip and magnetic film plane
respectively. The radial  component of the magnetic field of the
tip $H_r(R,d)$ at the edge $r = R$ can be expressed in terms of
the hypergeometric function:
\begin{equation}
H_r(R,d) = \pi \, M_t \, {R^3 \over d^3}\, _{2} F_{1} \left( {3 \over 2}, {3 \over 2}, 3,
-{4 R^2 \over d^2}\right),
\label{Hr}
\end{equation}
This value diverges logarithmically when the distance
between the tip and the magnetic film approaches zero:
\begin{displaymath}
H_r(R,d) \approx M_t \, \left\{ \begin{array}{ll}
 2\,\mbox{ln}\left({8 R \over e^2  d }\right) &  d \ll R \\
& \\
 {\pi \, R^3 / d^3}  &  d \gg R
\end{array}
\right\}
\label{Hr(R)}
\end{displaymath}

In a special case when the distance between the film and the tip
$d = 0$ the components of the magnetic field are:
\begin{eqnarray}
H_z(r) & = & - 2 \, \pi M_t \,\Theta[R - r],  \\
H_r(r) & = & \pi M_t \, \left({r \over
R}\right)^{\sigma}\, _{2} F_{1} \left({1 \over 2}, {3 \over 2}, 2, \left({r \over
R}\right)^{2 \sigma}\right),
\label{hzrd0}
\end{eqnarray}
where $\sigma = \mbox{sign}(R-r)$ and $\Theta$ is the Heaviside
function. The expression (\ref{hzrd0}) logarithmically diverges at
the distances close to the tip's edge $r = R$
\begin{equation}
H_r(r) \approx   2 \, M_t \mbox{ln}\left(\ {8 R \over e^2 |r-R|} \right)
\label{hrR}
\end{equation}
The components of magnetic field of the tip versus $r$ at
different $d$ are depicted in Fig. \ref{Htip}.

At large distances  $r \gg R$ the magnetic tip is equivalent to a
magnetic charge $Q = M_{t} \pi R^{2}$ and asymptotically the
magnetic field is Coulomb-like:
\begin{eqnarray}
H_z(r,d) & = &  - { \pi M_t R^2 d  \over (r^2 + d^2)^{3/2}}, \\
H_r(r,d)    & = & {\pi M_t R^2 r \over (r^2 + d^2)^{3/2}},
\label{Hr>R}
\end{eqnarray}

To make tip stronger by increasing magnetic charge of the tip $Q$
it is possible to increase the radius of the tip or its
magnetization. The first method is more effective.

\section{Static excitations in the presence of the tip}
\label{static}

\begin{figure}
\begin{center}
\vspace{10 mm}
\includegraphics[width=0.95\columnwidth,height=0.4\columnwidth]{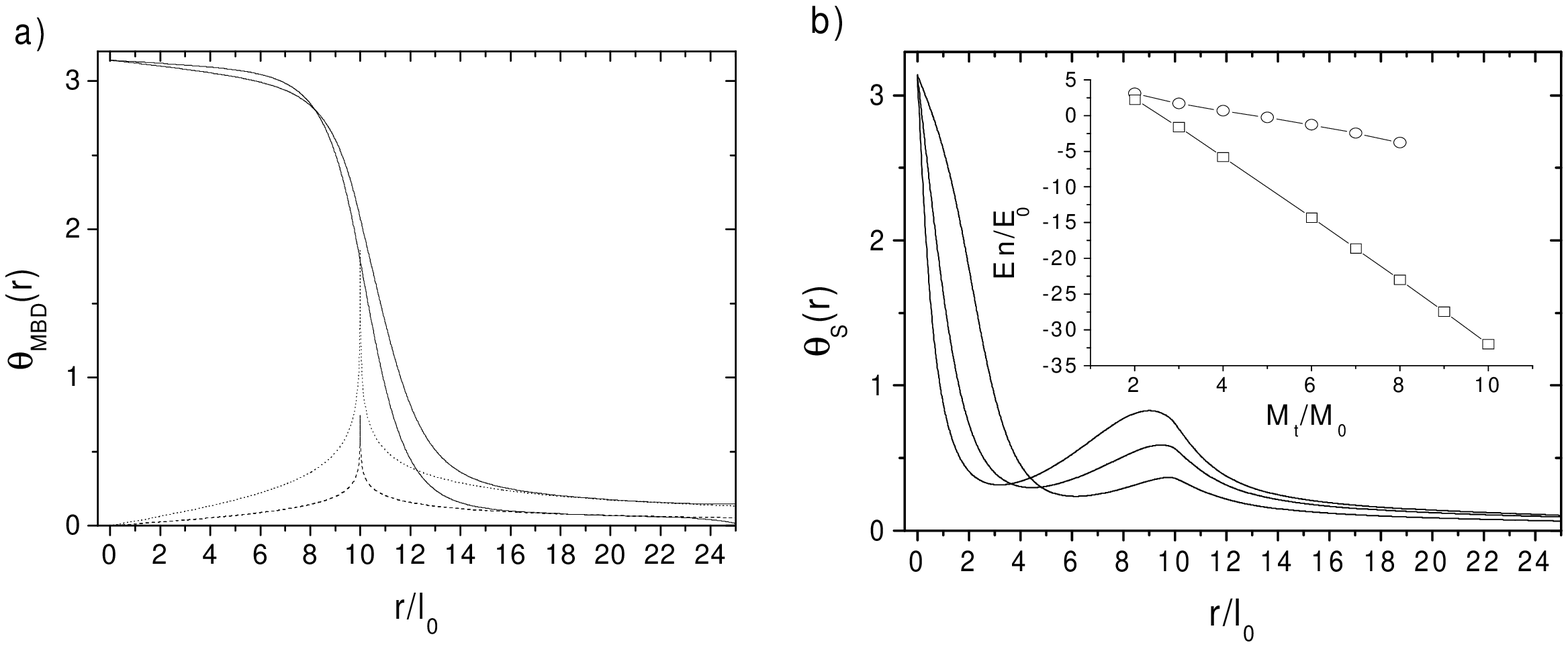}
\vspace{3 mm} \caption{ Profile of $\theta(r)$ at $d = 0$ and
different values of the ratio of the tip magnetization to that of
the film $M_t/M_0$. a) Profile of $\theta_{MBD}$ for $M_t/M_0
=4,10$. The radial component of magnetic field is shown on the
same figure. b) A Skyrmion-like solution strongly distorted by the
cusp of $H_r$ at the edge of the tip at $M_t/M_0  = 5, 7, 8$. The
energies of MBD $ \square $ and Skyrmion  $ \bigcirc $  versus
$M_t/M_0$ are shown in insert. $E_0 = 4 \pi \alpha h M_0^2$ is the
energy of the unperturbed Skyrmion. } \vspace{2 mm} \label{d0}
\end{center}
\end{figure}

In order to study the relaxation dynamics of the Skyrmion created
by  the magnetic tip it is necessary first to analyze static
solutions of the LLE for the magnetic film. A rough estimate of
the minimal tip radius stabilizing the MBD at very small distance
$d$ can be extracted from the balance of energy:
$R_0=\varepsilon_{dw}/(2\pi M_t M_0)$, where
$\varepsilon_{dw}= 2\sqrt{\alpha\beta} M_0^2$ is the linear tension of
the domain wall. For more accurate analysis the knowledge of the
magnetization and magnetic field distribution is necessary.

Several cylindrically symmetric soliton-like solutions of the LLE
have been found. They are the static Skyrmion solution
\cite{ivanov&all90,Abanov&Pokrovsky98}, the dynamic Skyrmion with
the spin precession \cite{abdullaev&all99} and the slowly
precessing magnetic bubble domain (MBD) \cite{sheka&all01}.

The precessing solitons can live longer than solitons without
precession, though they finally also collapse due to dissipation,
in accordance with the Derick-Hobart theorem
\cite{hobart63,derrick64}. The collapse does not proceed if the
Skyrmion configuration is stabilized by the higher order exchange
interaction. Stabilization also occurs in ferromagnets without
center of inversion \cite{ivanov&all90}.

Let us rewrite the LLE in terms of spherical coordinates $\theta$
and $\phi$, which determine the direction of magnetization:
\begin{widetext}
\begin{eqnarray}
{1 \over \omega_0} \sin \theta\, \dot{\theta}  =  l_0^2\, \mbox{div}\left(\sin^2\theta
\,\nabla \phi \left(1 + a^2 \kappa \,
(\nabla {\bf m})^2\right)\right) - {g H_r \over \omega_0}
\sin \theta \,\sin (\phi -  \varphi) - {\Gamma \over \omega_0} \dot{\phi} \, \sin^2 \theta,
\label{LL1}
\\
{1 \over \omega_0} \sin \theta \, \dot{\phi}  = l_0^2\,\left(
- \nabla^2 \theta + (\nabla\phi)^2 \sin\theta \cos\theta \right) (1 + a^2 \kappa (\nabla {\bf m})^2 )
- l_0^2\, a^2 \, \kappa \,  \mbox{div}((\nabla {\bf m})^2) +  \nonumber  \\
  \sin\theta\,
\cos\theta +
\, {g H_z\over \omega_0} \, \sin \theta - {g  H_r \over \omega_0}\, \cos \theta \,
\cos( \phi - \varphi) + {\Gamma \over \omega_0} \dot{\theta},
%}
\label{LL2}
\end{eqnarray}
where $(\nabla {\bf m})^2 = (\nabla \theta)^2 + \sin^2 \theta
\,(\nabla \phi)^2$, $\varphi$ is the azymuthal angle in the film
plane; $l_0 = \sqrt{\alpha/\beta} $ is the magnetic length and
$\omega_0 = \beta g M_0$ is the ferromagnetic resonance frequency.
In the static solution $\phi = \varphi$ and $\theta(r)$ obeys an
ordinary differential equation:
\begin{eqnarray}
{d^2 \theta \over d r^2} + {1 \over r}{d \theta \over dr} - \sin
\theta \cos\theta \left( {1 \over l_0^2 } + {1 \over r^2} \right)
+\nonumber\\
a^2 \kappa \left( {1 \over r}{d \over dr} \left( r\,(\nabla {\bf
m})^2 {d \theta \over dr} \right) - {\sin \theta \cos \theta \over
r^2  }(\nabla {\bf m})^2 \right) - h_z \sin\theta + h_r \cos
\theta  = 0, \label{theta}
\end{eqnarray}
where $h_{z, r} = H_{z, r}/\alpha M_0$. Equation (\ref{theta})
should be solved with the following boundary conditions:
$\theta(0) = \pi$, $\theta(\infty) = 0$.
\end{widetext}

One of the most powerful tools to solve the nonlinear ordinary
differential equations is the fitting method. The heart of this
method is the matching of a numerically derived solution to a
known asymptotics at small and large distances from the center of
nonlinear excitation by varying the shooting parameter.

The key problem in this method is to find a proper trial function,
with known asymptotics at $r \rightarrow 0$ and $r \rightarrow
\infty$ and fits it to the asymptotics of the real solution. We
solve LLE for $\theta(r)$ (\ref{theta}) by the fitting method. We
choose the trial function $\theta(r)$ in the following form:
\begin{equation}
\tan {\theta^{(f)} \over 2 } = {R^{(f)}\over r}\exp\left[-{(r-R^{(f)})\over l_0}\right].
\label{trial}
\end{equation}
The trial function (\ref{trial}) has two advantages. First, it
interpolates between two principal magnetic textures: the MBD and
the Skyrmion by varying the fitting parameter $R^{(f)}$, and
second, it has a valid asymptotic at $r \rightarrow 0$. Being
distorted by the radial component of the tip magnetic field $H_r$,
the radial component of the magnetization acquires an unusual
power asymptotics $\sim 1/r^2$. This behavior is more pronounced
when the tip is close to the magnetic film ($d \ll R$). In this
case the radial component of the tip magnetic field cannot be
treated as a perturbation, since it causes a strong distortion of
the Skyrmion. Small and large  values of $R_f$ correspond to the
Skyrmion solution (\ref{skyrm}) and the MBD, respectively. We
match the numerical solutions of LLE with the known asymptotic at
infinity.

Let us now consider a special case when the distance between the
surface of the and the tip $d$ is zero. The angle $\theta (r)$ at
large distances decreases as $1/r^2$, similarly to $h_r$. At small
distances, the solution for $\theta$ has the Skyrmion asymptotic.
We have solved numerically equation  (\ref{theta}) for $\theta$
with the shooting method for different $\mu = M_{t}/M_{0}$. For
each value of $\mu$ we have found two different values of the
fitting parameter $R_f$ satisfying the matching condition. We
treat the solution with larger $R_f$ as the MBD, whereas the
solution with smaller $R_f$ is interpreted as the Skyrmion-like
solution. Profiles of $\theta(r)$ for different values of $\mu$
are presented in Fig. \ref{d0}. The cusp of $h_r$ component of the
magnetic field of the tip produces the noticeable distortion of
Skyrmion solution in the region of the tip's edge. This distortion
grows with $\mu$ increasing as it is seen from Fig. \ref{d0}.
Strongly distorted Skyrmion-like excitation is the second solution
of Landau-Lifshitz equation, but its energy is larger than that of
the MBD. Thus, the magnetic tip nucleates MBD at static
conditions. Fig. \ref{d0} illustrates that the radius of the
generated MBD equals roughly to the radius of the tip. This result
can be obtained banalytically by the direct minimization of
energy.

\begin{figure}
\begin{center}
\vspace{10 mm}
\includegraphics[width=0.8\columnwidth,height=0.7\columnwidth]{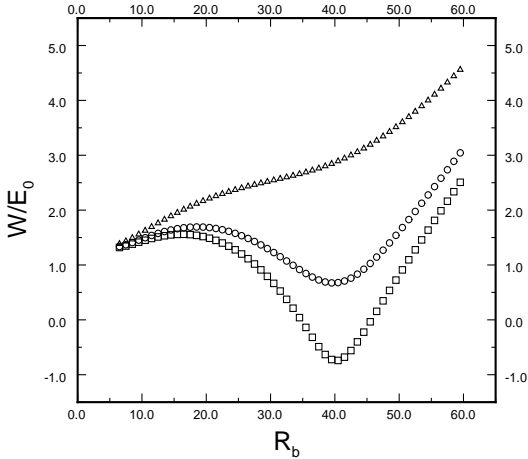}
%\vspace{5 mm}
\caption{The normalized energy of the MBD $W/E_0$ (\ref{freeen}) as a
function  of MBD radius
$R_b$ at different distances between the tip and film.
%% $\vartriangle$, $\square$, $\circle$
%% $\vartriangle, \square, \circ$
Curves with $\square$, $\bigcirc$ and $\bigtriangleup$ correspond to
$d/R$ = 0.1, 0.2, 0.5 respectively. The ratio of the radius of the tip to
the magnetic lenth $R/l_0 = 6$ and $\mu = 5$.
}
\label{bubenergy}
\end{center}
\end{figure}

In order to calculate the tip-film interaction energy we
substitute the expressions (\ref{HzHr}) for $H_z$ and $H_r$ and
equation (\ref{MBD}) into (\ref{H}). Using approximate relations:
$\mbox{sech}^2\left((r-R_b) /l_0\right) = 2\, l_0 \,
\delta(r-R_b)$ and $\mbox{sech}\left((r-R_b) /l_0\right) = \pi \,
l_0 \delta(r-R_b)$, we find
\begin{equation}
w_Z \equiv {W_Z \over E_0} = - \int_0^{\infty} k(q) \,
 J_1(q R)\, J_1(q \, R_b) e^{-q \, d} \, dq,
\label{Wz}
\end{equation}
where $k(q) = {2 \, \pi \mu R \, R_b \left( {1/q} + {\pi \, l_0
/2} \right) / (\beta \, l_0^2)} $, $E_0$ is the unperturbed
Skyrmion energy: $E_0 = 4 \pi \alpha M_0^2$; and $\mu = M_t/M_0$.
The total energy of the MBD $w$ in units $E_0$ is represented by
the sum of 4 terms
\begin{equation}
 w =  w_{E} + w_{a}  + w_{Z} + w_{D},
\label{freeen}
\end{equation}
where  $w_{E}$, $w_{a}$, $w_{Z}$ and $w_{D}$ are the exchange, the
anisotropy, Zeeman and demagnetization energies, respectively. The
partial sum $w_{E} + w_{a}$ can be represented by an interpolation
formula:
\begin{equation}
 w_{E} + w_{a} =  {R_b \over l_0} + 0.5 \,{l_0 \over R_b},
\label{exchanis}
\end{equation}
valid with good precision in a broad interval of the ratio
$R_b/l_0>2$. The first term in equation (\ref{exchanis}) is the
linear tension of the domain wall, the second is the correction to
it due nonzero topological charge of the MBD. 
The Zeeman energy is given by eq. (\ref{Wz}). The
demagnetizing energy $w_D$ for an ultrathin magnetic film ($h \ll
l_0$) slightly renormalizes the exchange energy and can be
neglected in the first approximation. This approximation is
correct provided the radius of the MBD is much larger than the
thickness of the film.

The energy of MBD versus its radius  $R_b$ for different $d$ is
shown in Fig. \ref{bubenergy}. If $d > d_c = 0.385\, R$, the
energy minimum at $R_b=R$ disappears; the tip cannot confine MBD
anymore and the latter shrinks.

Accepting $R_b = R$, we find from Eq. (\ref{freeen}) the range of
parameters, in which the nucleation of the MBD is favorable:
\begin{equation}
{R \over l_0} > {1 - 0.5 \, \pi \, \xi \, g(p) \over \xi \, f(p)},
\label{phasediag1}
\end{equation}
where $\xi = \pi \mu/\beta$,
\begin{equation} g(p) = \,_{2}F_{1}
\left( {3 \over 2}, {3 \over 2}, 3, -{4 \over p^2}\right)/p^3,
\label{g}
\end{equation}
\begin{equation}
f(p) = 1 - {2 p \over \pi \sqrt{k}}\left[ K(k) - E(k) \right]
\label{f}
\end{equation}
$p = d/R$, $k = (1 + p^2/4)^{-1}$; $K(k)$ and E(k) are the
complete elliptic integrals of the first and second kind,
respectively.

The total energy of the MBD has the following form
\begin{equation}
 w = \zeta + 0.5/\zeta - \xi \, \zeta^2 \, f(p) \, \cos \psi - 0.5 \pi \, \xi \, \zeta \, g(p),
\label{toten}
\end{equation}
where $\zeta = R_b/l_0$ and $\psi$ is the global rotation angle.
In the equlibrium state $\psi = 0$.

At a fixed radius of the tip $R$, the bubble nucleates when the
distance between the tip and the film becomes less than some
critical value $d_c$.

\begin{figure}
\begin{center}
\vspace{10 mm}
%\vspace{-30 mm}
\includegraphics[width=0.9\columnwidth,height=0.7\columnwidth]{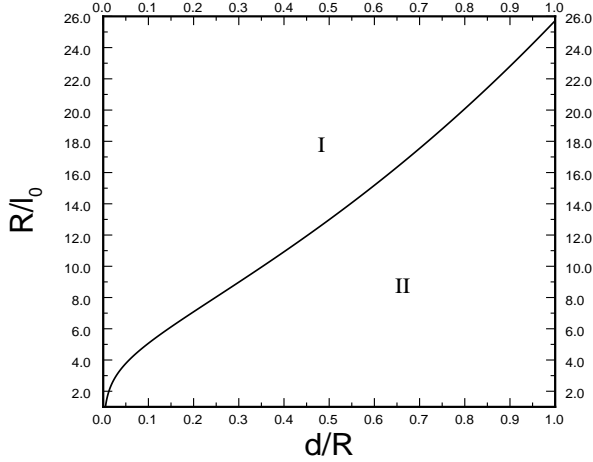}
\caption{ The phase diagram showing the range of parameters in
which the nucleation of the MBD is energy favorable (region I).
$d$ is the distance between the tip and the surface of the film;
$R$ is the tip radius, $\mu = 5$, $\beta = 100$. }
\label{phasediag}
%\vspace{5 mm}
\end{center}
\end{figure}

\section{Discrete spin model with the dipolar interaction}
\label{discrete} In this section we study the dynamics of the
bubble creation and its transformation into a stable  Skyrmion
solving numerically time-dependent Landau-Lifshitz equation on a
discrete lattice. We employ the model of a classical
two-dimensional Heisenberg ferromagnet with the nearest, next
nearest (NNN), and next to next nearest neighbor (NNNN)
interaction and dipolar interaction. The corresponding Hamiltonian
reads:
\begin{eqnarray}
{\cal H} &=& - J\sum_{{\bf n}, \, {\bf a}} {\bf
S}_{\bf n}\cdot {\bf S}_{{\bf n} + {\bf a} } - 0.5 \lambda \sum_{\bf n} (S_{\bf n}^z)^2
 \nonumber \\ & &
+ D \sum_{{\bf n}, {\bf n'}}\left(
{ {\bf
S}_{\bf n}\cdot {\bf S}_{\bf n' } \over {\bf r}_{{\bf n} {\bf n'}}^3 }
- 3 {
({\bf
S}_{\bf n}\cdot {\bf r}_{{\bf n} {\bf n'}}) ({\bf
S}_{\bf n'}\cdot {\bf r}_{{\bf n} {\bf n'}})
\over {\bf r}_{{\bf n} {\bf n'}}^5
}
\right) \nonumber \\ & &
- J'\sum_{{\bf n}, \, {\bf b}} {\bf
S}_{\bf n}\cdot {\bf S}_{{\bf n} + {\bf b} }
+ J"\sum_{{\bf n}, \, {\bf a}} {\bf
S}_{\bf n}\cdot {\bf S}_{{\bf n} + 2 {\bf a} }
\nonumber \\ & &
- \sum_{\bf n}{\bf S}_{\bf n}\cdot
{\bf H},
\label{disH}
\end{eqnarray}
where the summations run over all spin sites ${\bf n}$, its
nearest neighbors ${\bf a}$, NNN along diagonals ${\bf b}$ and
NNNN $2 {\bf a}$. We assume the ferromagnetic exchange $J$ between
NN, antiferromagnetic exchange  $J"$ between NNNN; the sign of NNN
exchange $J'$ is not fixed; $\lambda$  denotes easy axis
anisotropy constant and $D$ is the dipolar long range coupling
constant. Both exchange interactions $J'$ and $J"$ mimic the real
RKKY interaction in metallic ferromagnets mediated by conduction
electrons. In usual magnets the dimensionless ratio $D/(Ja^3)$ is
of the order $10^{-4} - 10^{-3}$. Fourier transform of the
oscillatory RKKY interaction has the form $F(q) = -(\nu/4) f(q/2
k_{F})$, where $\nu = p_F \, m/(\pi^2 \, \hbar^3 )$ is the density
of states and
\begin{equation}
f(x) = {1 \over 2} \left( 1 + {1-x^2 \over 2 x} \mbox{ln} \left|
{1+x \over 1-x} \right| \right)
\end{equation}
is the Lindhardt function, whose expansion into the Tailor series
over powers  of $x$ is
\begin{equation}
f(x) = 1 - {x^2 \over 3} - {x^4 \over 15}  - \ldots -{x^{2 n} \over (2 n-1) \cdot (2 n+1)} - \ldots
\label{lindhardt}
\end{equation}
Since both coefficients at $x^2$ and $x^4$, have the same sign,
the ferromagnetic state provided by this interaction  satisfies
the condition of stability (see Section \ref{model}).

In the continuous limit the Hamiltonian (\ref{disH}) takes a
following form
\begin{equation}
{\cal H}= {1 \over 2}\,\int \tilde{J} (\nabla {\bf S})^{2}\,d^2 x + {\cal H}_{1} +  {\cal H}_D,
\label{contH}
\end{equation}
where $\tilde{J} = J + 2J' - 4J"$, ${\cal H}_D$ is the
demagnetization field energy. The perturbative part of the
Hamiltonian ${\cal H}_{1}$ is
\begin{equation}
{\cal H}_{1} = \int  \left( { J" a^4  \over 2}(\nabla^2 {\bf S})^{2} +
{\lambda \over 2}(1 - S_z^2) - {\bf h} \,{\bf S}\right)
\,{d^2 x \over a^2} ,
\label{contH1}
\end{equation}
where ${\bf h} = g \mu_B {\bf H}$ is the scaled magnetic field.
The Skyrmion is stable in the continuous model if both $\tilde{J}$
and $J^{\prime\prime}$ are positive. The discreteness changes the
topology and make the continuous decay of Skyrmion possible. To
study the discreteness effect in continuous model Haldane has
proposed  to remove one plaquette in the center of the Skyrmion
\cite{haldane88}. An explicit calculation of the Skyrmion decay by
this mechanism was performed in the work
\cite{Abanov&Pokrovsky98}. The Skyrmion is metastable in a
discrete lattice if in a configuration with only one reversed spin
its rotation to the parallel orientation requires overcoming a
potential barrier. This requirement is satisfied in a square
lattice if $J^{\prime\prime}>0.15(J+2J^{\prime})$. Though
important for simulation on a discrete lattice, this effect does
not play any role for real itinerant magnets, in which spin
density is continuous.
%\begin{equation} 0 < 0.15 \, (J + 2 \, J')
%< J" < 0.25 \, (J +  2 \, J'). \label{crit}
%\end{equation}
As we have shown earlier, the Skyrmion stability requires
$J^{\prime\prime}$ to be positive, whereas the sign of
$J^{\prime}$ remains indefinite.

\section{From bubble to skyrmion}
\label{relax}

To study the dynamic transformation of the MBD into the Skyrmion
we solved the time-dependent LLE by the Runge-Kutta method for a
sample with free boundaries. The sample had a circular shape and
contained 11308 spins located at the sites of a square lattice. We
choose time step 10$^{-3} \, \omega_0^{-1}$ for good convergence.
For dynamical simulations we generated initially the MBD imitating
a strong tip in a close vicinity of the magnetic film. The radius
of the generated MBD is approximately equal to the radius of the
tip (the difference between radii of the tip and the MBD is below
5 \%).

\begin{figure}
\begin{center}
\vspace{10 mm}
\includegraphics[width=0.9\columnwidth,height=0.9\columnwidth]{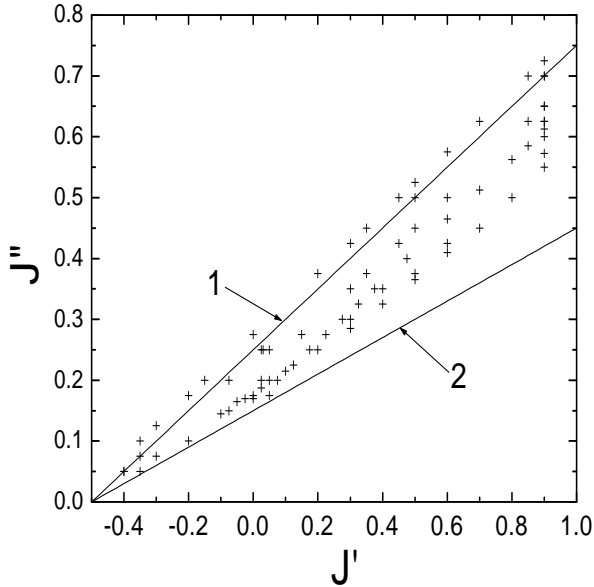}
%\vspace{18 mm}
\vspace{5 mm} 
\caption{ Simulation of spin dynamics at different
$J'$ and $J''$: crosses correspond to a stable Skyrmion, curve 1
corresponds to $J'' = 0.25 + 0.5 \,J'$, curve 2 to $J'' = 0.15 +
0.3 \,J'$. Simulations are in a good agreement with the criterion
of stability against long-wave spin perturbations.}
\label{jjstab}
\end{center}
\end{figure}

Once the bubble was generated, we started to move away the tip
uniformly ($d(t) = d_0 + v t$). The MBD gradually transformsed
into a stable Skyrmion or collapsed depending on the exchange
parameters $J'$ and $J"$. The range of the exchange parameters
corresponding to the Skyrmion stability is represented in Fig.
\ref{jjstab}.  The Skyrmion stability criterion 
%(\ref{crit}) 
is in
good agreement with the numerical data.

\begin{figure}[ht]
\centering
 \includegraphics[width=0.9\columnwidth,height=0.7\columnwidth]{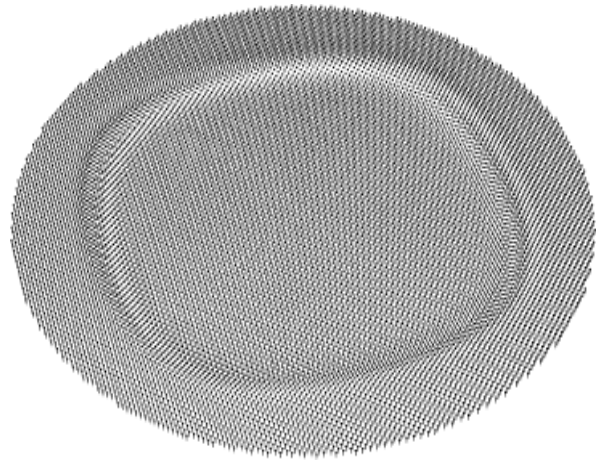}
\caption{Magnetic bubble generated by numerical simulations. A
small square distortion of axial symmetry is caused by symmetry of
the lattice. This effect becomes pronounced during the dynamical
transformation of the bubble into the Skyrmion. The ratio of the
distance between the tip and film is $d/R = 7/4$. }
 \label{bubble}
\end{figure}

The spin structure of the MBD is shown in Fig. \ref{bubble}. The
shape of the bubble is slightly distorted during the collapse.
This distortion owes to the effect of the square lattice.

We also studied the difference between continuous and discrete
approximation by numerical simulations with different anisotropy
constants $\lambda$. The discrete models approaches the continuous
approximation when anisotropy constant $\lambda$ decreases and the
domain wall width $l_0$ increases (See Table \ref{table1}).

\begin{table}
\caption{Normalized parameters of the soliton versus anisotropy
constant $\lambda$ derived from numerical simulations. All
parameters are normalized to ferromagnetic exchange interaction
$J$. Antiferromagnetic exchange interaction constant $J" = 0.2$
and $J' = 0$.}
\begin{tabular}{l|c|c|c|c|c|c|c}
%\begin{center}
\hline
\hline

Anisotropy & & & & & & & \\
constant,
$\lambda$  & 0.001 &  0.0025 &  0.005 & 0.01  & 0.02  & 0.03  & 0.04 \\
\hline
Effective  & & & & & & & \\
AFM  & 0.19     & 0.186      &  0.181    & 0.175  & 0.166  & 0.16   & 0.155 \\
exchange, $J"_{eff}$  & & & & & & & \\
\hline
Skyrmion  & & & & & & & \\
 radius, $R_s$       &  4.28   &   3.8      &  3.22    &  2.77  &   2.36   &  2.13  & 1.97 \\
\hline
Domain & & & & & & & \\
wall-width, $l_0$       &   10.39   &     8.78    &    7.88    &   7.36 &   7.08  &  6.79  & 6.53\\
\hline
\hline
%\end{center}
\end{tabular}
\label{table1}
\end{table}

The process of transformation passes three stages. In the first
stage the bubble domain is controlled by the removing tip until it
reaches the distance $d\approx 2R$. During the second stage the MBD
shrinks under the action of the linear tension. When its radius
reaches the value $R\sim l_0$, it can not be more described as the
MBD and acquires the profile of a Skyrmion. A rough interpolation
formula for the transformation time $T(R)$ following from
semi-intuitive consideration reads:
\begin{equation}
 T \approx {2 R \over v} + {R^2 \over l_0 \, v_m \, \Gamma} + {1 \over \omega_0 \, \Gamma}.
 \label{totaltime}
 \end{equation}
In this equation the first term is the time during which the tip
reaches the distance $d=2R$ from the film. At larger distance the
interaction between the tip and the film is negligibly small. The
second term corresponds to the shrinking of the MBD under the
action of the Laplace force from the domain wall. The Laplace
force is equal to $F_L=\sigma /R$, where $\sigma = 2 \sqrt{\tilde{J} \lambda}/a$ 
is the linear tension. The local velocity of the domain
wall is antiparallel to the radius and equal to $\eta F_L$, where
$\eta =\Gamma  v_m/(4 \lambda)$ is the magnetic viscosity. The time
$T_1(R)$ necessary to decrease significantly the radius is equal
to integral
$$T_1(R)=\int_0^R\frac{r \, dr}{\eta \, \sigma}.$$
This gives exactly the second term in equation (\ref{totaltime}).
Finally, the third term is the relaxation time for the Skyrmion
derived in \cite{Abanov&Pokrovsky98}. This simple interpolation
fits the simulation data with very good accuracy.

A more sophisticated description of the transformation process can
be obtained by substitution of the interpolation between the MBD
and Skyrmion given by equation (\ref{trial}) into the Hamiltonian
(\ref{H})and integrate it over all variables. The resulting
expression is the Hamiltonian for slow variables $R$ and $\psi$.
The dimensionless Lagrangian density $l$ can be constructed given the
Hamiltonian density $w$ by the following ansatz:
\begin{equation}
l = 0.5 \, \zeta^2 \, \frac{\partial\psi}{\partial \tau} - w, 
\label{LMBD}
\end{equation}
where $\zeta = R_e/l_0$, $\tau = \omega_0 t$ and $R_e$ is the radius of topological
excitation.
To incorporate the dissipation it is possible to use the Rayleigh
dissipative function $\cal{R}$. In dimensionless variables it
reads ${\cal R}=\frac{1}{2}\Gamma \, \zeta \,(\zeta_{\tau}^2+\psi_{\tau}^2).$ 
We will not present the equations of motion here to avoid lengthy
formulas with no essentially new results. They can be useful when
a more detailed description of the process will be in order.

\begin{figure}[ht]
\vspace{20 mm}
\centering
 \includegraphics[width=0.9\columnwidth,height=0.7\columnwidth]{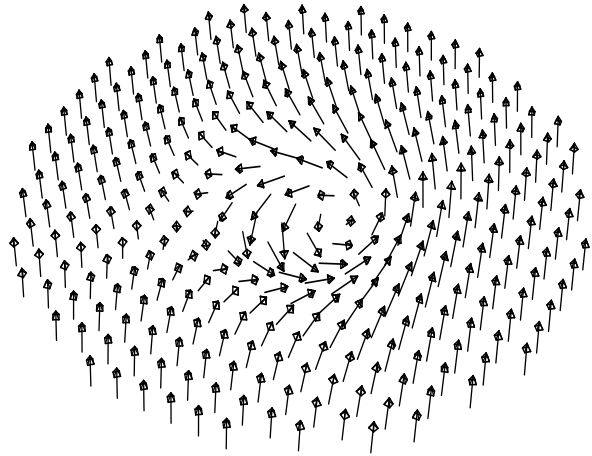}
\caption{The Stable Skyrmion spin configuration generated from
numerical simulations. The anisotropy constant is $\lambda =
0.04$, the exchange interaction constants are $J = 1$, $J' = 0.6$,
$J" = 0.5$. }
\label{skyrmion}
\end{figure}

For real magnetic film with $l_0 \approx 10 \, nm$, tip with
radius $R = 50 \, nm$, damping constant $\Gamma = 0.1$ and tip's
velocity $v = 1 \, nm/ns$ transformation time is $T \approx 125 \,
ns$. The equilibrium size of the Skyrmion is
 \begin{equation}
 R_0 = a \left(\, {8 J" \over 3 \lambda L}\right)^{1/4} \approx 1 \div 3 \, nm,
 \label{R0}
 \end{equation}
where $L = \ln\left[(3 \lambda/8 J")^{1/4} l_0 \right]$ and $a$
is the lattice spacing. Skyrmion spin structure obtained by
numerical simulations of time-dependent Landau-Lifshitz Equation
is represented in Fig. \ref{skyrmion}.

An effective magnetic field corresponding to the exchange
interaction $\tilde{J} $ is $ H_{J} = \tilde{J}\,a/\mu_B$. For
$Fe$ it is about 89 T and for $Co$ it is 112 T. For existing
strong magnetic tips real value of created magnetostatic field is
about 1 T. Therefore we consider the ratio $H/2 \pi  \tilde{J} =
0.001$. For iron film and tip with radius 50 nm minimum magnetic
field of the tip $H_{tip} \approx  28 \, G $.

\section{Conclusions and Discussion}
\label{conclusion}

The main result of this work is the numerical evidence that the
bubble domain of the 50 nm size initiated by a strong magnetic tip
in a thin magnetic film shrinks into a very small, few nanometers
metastable Skyrmion. For this purpose we employed the MC and the
Runge-Kutta methods on the square lattice. The static Skyrmion and
the magnetic bubble domain were found numerically as solutions of
the static Landau-Lifshitz equation.

The role of the magnetic tip is two-fold. Its magnetic field
stabilizes the bubble domain, which otherwise is unstable in a
thin film. Being removed it does not confine the MBD anymore and
the latter starts to shrink conserving its topological charge. The
Skyrmion is the final result of this shrinking. Thus, the position
of the Skyrmion is determined with high accuracy by the tip. This
is important for the experimental observation of the Skyrmion,
since it is very small and its magnetic field is strongly
localized as it was established by our calculations. On the other
hand, strong localization of the Skyrmion magnetic field makes
possible to read the information stored by Skyrmions if the
distances between them and between them and empty spaces are more
than $3 R_s$, where $R_s$ is Skyrmion radius. Such a distance is
sufficient to ensure that Skyrmions do not perturb the
magnetization around.

Limitations on exchange constants accepted in our work stem from
the requirement of the Skyrmion stability. One of them is
associated with the discrete character of the model and is not
essential for a real itinerant magnet with continuous distribution
of magnetization. Another one occurs in continuous model: the part
of energy proportional to the fourth power of the gradient must be
positive. Being imposed onto a lattice model it leads to an
inequality, which provides sufficiently large antiferromagnetic
exchange between the NNNN. In real itinerant ferromagnets the
natural RKKY interaction ensures the stability.

Thus, the Skyrmion stability makes it is possible to create it by
circular magnetic tip and to observe it by spin-polarized scanning
tunnelling microscopy having a nanometer resolution. It is not yet
clear how the creation of the next Skyrmion by lateral scanning
tip will influence an already existing Skyrmion. When this problem
will be solved, the Skyrmion arrays could be a promising candidate
for the magnetic memory of unprecedented high density above 10
TBit/in$^2$.

\section{acknowledgments}

This work has been supported by NSF under the grants DMR0072115
and DMR0103455, by DOE under the grant FG03-96ER45598 and by the
Telecommunications and Information Task Force at Texas A\&M
University.

%\newpage

%\vspace{15 mm}
%\newpage

%\vskip35mm

%\newpage

%\newpage
%\end{thebibliography}

\end{document}